%% file: main.tex
\documentclass[preprint,journal]{vgtc}            % preprint (journal style)
\onlineid{0}

\preprinttext{Preprint version for arXiv submission.}

\vgtccategory{Research}

\vgtcpapertype{theory/model}

\title{A Directed Graph Model and Experimental Framework for \\
Design and Study of Time-Dependent Text Visualisation}
\author{%
  \authororcid{Songhai Fan}{0009-0004-8456-2613},
  Simon Angus\textsuperscript{\textsection},
  Tim Dwyer\textsuperscript{\textbardbl},
  Ying Yang\textsuperscript{\textdagger},
  Sarah Goodwin\textsuperscript{\textdaggerdbl},
  Helen Purchase\textsuperscript{\textparagraph}
}

\authorfooter{
  \item
  Songhai Fan is with Monash University.
  E-mail: songhai.fan@monash.edu.
  \item
  Simon Angus is with Monash University.
  E-mail: simon.angus@monash.edu.
  \item
  Tim Dwyer is with Monash University.
  E-mail: tim.dwyer@monash.edu.
  \item
  Ying Yang is with Monash University.
  E-mail: ying.yang@monash.edu.
  \item
  Sarah Goodwin is with Monash University.
  E-mail: sarah.goodwin@monash.edu.
  \item
  Helen Purchase is with Monash University.
  E-mail: helen.purchase@monash.edu.
}

\abstract{
Exponential growth in the quantity of digital news, social media, and other textual sources makes it difficult for humans to keep up with rapidly evolving narratives about world events. Various visualisation techniques have been touted to help people to understand such discourse by exposing relationships between texts (such as news articles) as topics and themes evolve over time. Arguably, the understandability of such visualisations hinges on the assumption that people will be able to easily interpret the relationships in such visual network structures. To test this assumption, we begin by defining an abstract model of time-dependent text visualisation based on directed graph structures. From this model we distill motifs that capture the set of possible ways that texts can be linked across changes in time. We also develop a controlled synthetic text generation methodology that leverages the power of modern LLMs to create fictional, yet structured sets of time-dependent texts that fit each of our patterns. Therefore, we create a clean user study environment ($n=30$) for participants to identify patterns that best represent a given set of synthetic articles. We find that it is a challenging task for the user to identify and recover the predefined motif. We analyse qualitative data to map an unexpectedly rich variety of user rationales when divergences from expected interpretation occur. A deeper analysis also points to unexpected complexities inherent in the formation of synthetic datasets with LLMs that undermine the study control in some cases. Furthermore, analysis of individual decision-making in our study hints at a future where text discourse visualisation may need to dispense with a `one-size-fits-all' approach and, instead, should be more adaptable to the specific user who is exploring the visualisation in front of them.

}


\teaser{
  \centering
  \includegraphics[width=\linewidth]{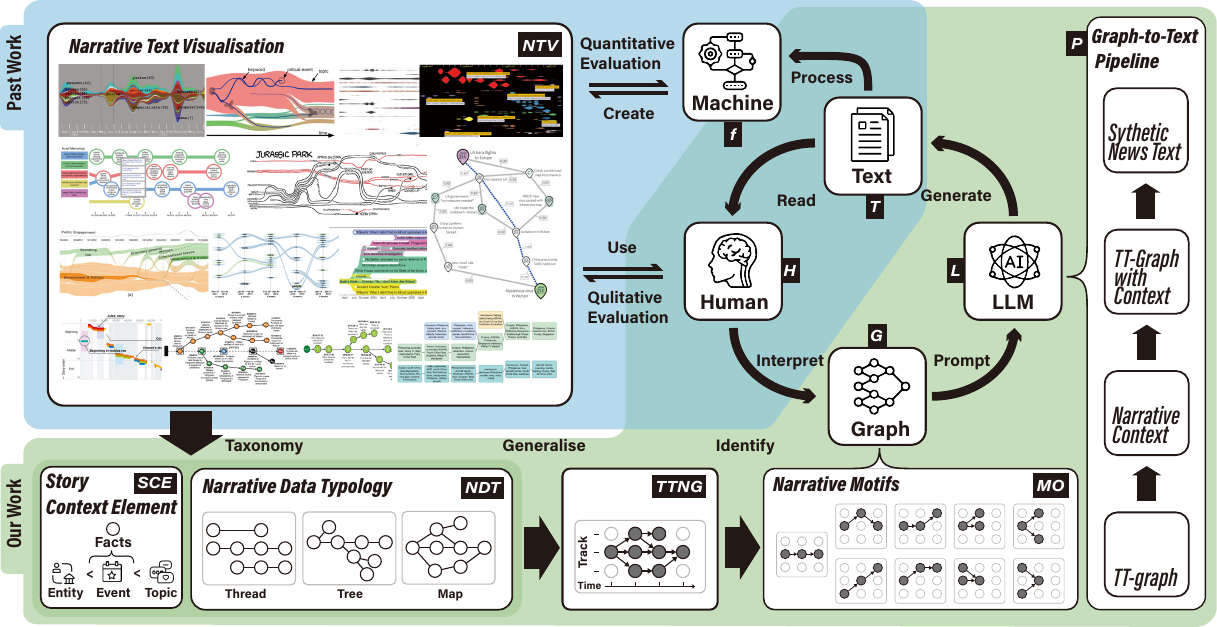}
  \caption{%
  	Overview of workflow model for evaluating human understanding of text.%
  }
  \label{fig:teaser}
}

\graphicspath{{figs/}{figures/}{pictures/}{images/}{./}} % where to search for the images
\usepackage{mathptmx}                  % use matching math font

\usepackage[utf8]{inputenc}            % Allow UTF-8 input encoding
\usepackage{amsmath}                   % Enhanced math typesetting
\usepackage{bm} %bold math

\usepackage{amsfonts}                  % Additional mathematical fonts

\usepackage{graphicx}                  % Include and manipulate graphics

\usepackage{amsthm}                    % Define environments for theorems, definitions, examples, etc., which provide a standardized format for these elements.

\usepackage{tabularx}
\usepackage{algorithm}
\usepackage{algpseudocode}

\usepackage{wrapfig}
\usepackage{float}
\usepackage{dblfloatfix} % Fixes the placement of double-column floats

\usepackage{multirow}

\usepackage{listings} % Include the package for listings
\usepackage{enumitem} % Keep list spacing consistent and compact
\setlist[itemize]{leftmargin=*,itemsep=1pt,topsep=2pt,parsep=0pt,partopsep=0pt}
\setlist[enumerate]{leftmargin=*,itemsep=1pt,topsep=2pt,parsep=0pt,partopsep=0pt}
\setlist[description]{itemsep=1pt,topsep=2pt,parsep=0pt,partopsep=0pt}
\newcommand{\teaserLink}[1]{%
  \colorbox{black}{\textcolor{white}{#1}}%
}

\begin{document}

%%%%%%%%%%%%%%%%%%%%%%%%%%%%%%%%%%%%%%%%%%%%%%%%%%%%%%%%%%%%%%%%
%%%%%%%%%%%%%%%%%%%%%% START OF THE PAPER %%%%%%%%%%%%%%%%%%%%%%
%%%%%%%%%%%%%%%%%%%%%%%%%%%%%%%%%%%%%%%%%%%%%%%%%%%%%%%%%%%%%%%%

\input{inputs/01_introduction}

% Section 2: Taxonomy of Narrative Text Visualisation
\input{inputs/02_taxonomy/01_overview}

\input{inputs/02_taxonomy/02_elements}
\input{inputs/02_taxonomy/03_evaluation}

% Section 3: TTNG Model and Motifs
\input{inputs/03_model/01_overview}
\input{inputs/03_model/02_definition}
\input{inputs/03_model/03_motifs}

% Section 4: Graph-to-Text Pipeline
\input{inputs/04_pipeline/01_overview}
\input{inputs/04_pipeline/02_implementation}
\input{inputs/04_pipeline/03_validation}

% Section 5: User Study
\input{inputs/05_study/01_design}
\input{inputs/05_study/02_validation}
\input{inputs/05_study/03_procedure}

% Section 6: Results
\input{inputs/06_results/01_overview}
\input{inputs/06_results/02_quantitative}
\input{inputs/06_results/03_qualitative}
\input{inputs/06_results/04_suggestion}

% Section 7-8: Discussion and Conclusion
\input{inputs/07_discussion}
\input{inputs/08_conclusion}

\bibliographystyle{abbrv-doi-hyperref}
%\bibliographystyle{abbrv-doi-hyperref-narrow}
%\bibliographystyle{abbrv-doi}
%\bibliographystyle{abbrv-doi-narrow}
% \bibliography{template}
\bibliography{references,references_manual}
\clearpage
\onecolumn
\appendix % You can use the `hideappendix` class option to skip everything after \appendix
\input{additional}
\end{document}

%% file: inputs/01_introduction.tex
%% The ``\maketitle'' command must be the first command after the
%% ``\begin{document}'' command. It prepares and prints the title block.
%% the only exception to this rule is the \firstsection command
\firstsection{Introduction}
\maketitle
% \section{Introduction} %for journal use above \firstsection{..} instead
As online discourse related to world news grows exponentially, discerning meaningful patterns and trends becomes increasingly challenging for people awash in information. Visualisation has been proposed as a tool for helping readers understand complex and evolving narratives, exploiting human capacity for visual processing\cite{nualart-vilaplana_how_2014,alharbi_sos_2019}. Various visual metaphors have been proposed to represent time-dependent change in textual data, such as theme \textbf{rivers}\cite{havre_themeriver_2002}, storylines\cite{xu_summarizing_2013}, narrative \textbf{maps}\cite{keith_norambuena_narrative_2021}, story \textbf{forests}\cite{liu_story_2020} and others (see Section \ref{sec:background}). Researchers have focused on different properties when proposing designs for visualisation of time-dependent text data. Examples include evolution of topics\cite{zhou_survey_2017}, stance and sentiment\cite{kucher_sentiment_2019}, and timelines of key events\cite{brehmer_timelines_2017}.
However, despite this large variety of academic work, visualisation of time-dependent text such as narratives in news articles remains largely underutilised in ecological settings. This is particularly problematic when multiple narratives coexist and conflict over time, making interpretive comparison difficult.
To explore how people respond to visual representations of such narratives, in this paper we step back from specific approaches to time-dependent text visualisation to define a data model, instantiated as a directed graph, for the design and study of visualisation of changing narratives across sets of time-dependent texts.

Considering the information overload problem faced by people exposed to social media and other online information sources, we focus on datasets where each text conveys a statement of an event at a different point in time (the classic example being corpora of news articles). While many other domains produce time-dependent texts (e.g., speeches, scripts, blogs, and tweets), news articles are particularly convenient units of analysis: each article is typically a self-contained account of an event tied to a timestamp. Focusing on news therefore simplifies model design and evaluation, and the same principles can be extended to other domains once their domain-specific idiosyncrasies are addressed.
Formalising such a unified model can progress time-dependent text visualisation in the news domain in several ways.

First, as identified in a recent survey of computational understanding of text \cite{ranade_computational_2022}, there is a lack of a standardised framework for extracting narrative from text. The time-dependent text visualisation problem in the news domain could benefit from formal definitions. As noted in Section \ref{sec:background}, researchers have created valuable techniques for visualising narrative structures, though these are spread across disciplines. A unified framework could promote consistency and support the development of broad evaluation methods.

Second, validating proposed text visualisation techniques is still rare in the literature, presumably due to the high costs associated with collecting well-annotated or organic text datasets as ground truth for evaluation. Furthermore, while there are abundant texts available that describe real-world news, people (subjects) will bring their own knowledge of this news to any interpretive task \cite{zhao_evaluating_2022}, potentially contaminating user studies where in-memory news items are used. Datasets free of context and with precisely controlled structure are required to assess the effectiveness of different visualisation techniques reliably.

As illustrated in the green part of figure \autoref{fig:teaser}, our approach aims to address these challenges through a novel and structured methodology. Our work can be broken down into the following key contributions (\teaserLink{\textit{black text boxes}} correspond to labels in \autoref{fig:teaser}):

\noindent\teaserLink{$ NTV \longrightarrow SCEs,NDTs \longrightarrow TTNG $}

\noindent \textbf{Taxonomy and Formalisation for Time-Dependent Text Visualisation.} Through a review of prior work, we categorise key elements of narrative-oriented visualisation into Story Context Elements (SCEs) and Narrative Data Topologies (NDTs). We then formalise these elements in our Time-Track Narrative Graph (TTNG) model as a visualisation-oriented data model, providing a unified intermediate representation between texts and visual encodings, and between narrative specifications and text generation, for time-dependent text.

\noindent\teaserLink{$ G \longrightarrow L \longrightarrow T $}
\textbf{Low-Cost Synthetic Dataset Generation.}
We introduce a \textit{Graph-to-Text Pipeline}, a novel methodology to generate controlled, synthetic datasets at low cost. This methodology leverages state-of-the-art language models (LLMs) to create datasets that can be used for rigorous evaluation of time-dependent text visualisation techniques in the news domain.

\noindent\teaserLink{$ T \longrightarrow H \longrightarrow G $}
\textbf{Exploratory study of Text-to-Graph Visualisation.}
We present a study demonstrating the application of our formalism and a methodology involving generation of synthetic data. This study explores people's ability to relate simple texts to basic visualisation patterns. The findings provide early evidence and help inform our ongoing efforts to design visualisations of time-dependent news that support comprehension.

Our results indicate that the generation of synthetic datasets requires better control and validation of reliability. Predefined narrative structure, although validated by machine, is not well understood by human subjects, even with linear sequential narratives. Further, we observe significant variation in subject understanding and interpretation of visualisations. We recommend standardising a framework for visualising time-dependent text in the news domain to enable consistent research. Improving dataset control and validation will improve the reliability of the evaluation. Future work should focus on designing more interactive and customisable visualisations.

The paper is structured as follows. Section~\ref{sec:background} \teaserLink{$NTV$} presents a taxonomy of time-dependent text visualisation in the news domain derived from our review of prior work. Section~\ref{sec:tt_graph_model} \teaserLink{$TTNG$} formalises this taxonomy into the TTNG model and introduces narrative motifs \teaserLink{$MO$}. Section~\ref{sec:graph_to_text} \teaserLink{$P$} describes our Graph-to-Text pipeline for generating controlled datasets. Section~\ref{sec:experimental_design} details our user study design, and Section~\ref{sec:results} presents both quantitative and qualitative results. Finally, Sections~\ref{sec:discussion} and \ref{sec:conclusion} discuss limitations and future research directions.

%% file: inputs/02_taxonomy/01_overview.tex
\section{Related Work, Definitions and Taxonomy}
\label{sec:background}

\input{figs/taxonomy_table}

The scope of this work spans Natural Language Processing~(NLP), Information Visualisation~(InfoVis), Visual Analytics~(VA), and Human-Computer Interaction~(HCI) evaluation in text visualisation. Through the literature review, we identify and categorise recurring patterns in time-dependent text visualisation approaches in the news domain. While the taxonomy is framed to apply to time-dependent text more broadly, in this paper we emphasise news to ground definitions and evaluation. This review leads us to propose a high-level taxonomy with two key aspects (explained in the following section): \textbf{Story Context Elements (SCEs)} that represent the building blocks of news narratives, and \textbf{Narrative Data Topologies (NDTs)} that represent how these elements are connected.

%% file: figs/taxonomy_table.tex
\begin{table*}[htp]
\centering
\caption{Taxonomy of Narrative Visualisation Techniques: This table categorises various time-dependent text visualisation techniques based on three dimensions: the domain of text data (e.g., news), Story Context Elements (entity, event, topic), Narrative Data Topologies (thread, tree, map), and evaluation methods (whether a case study or user study is involved, and whether the approach is qualitative (Ql) or quantitative (Qt)). Row numbers correspond to labelled example images in \autoref{fig:teaser}. Ticks indicate the presence of specific characteristics within each technique.}
\label{tab:narrative_taxonomy}
\vskip 5pt 
\resizebox{\textwidth}{!}{%
\begin{tabular}{ll|ccc|ccc|ccc}
  \multicolumn{1}{c|}{\teaserLink{$NTVs$}} &
  \multicolumn{1}{c|}{\textbf{Domain}} &
  \multicolumn{3}{c|}{\textbf{Story Context Element}} &
  \multicolumn{3}{c|}{\textbf{Narrative Data Topology}} &
  \multicolumn{3}{c}{\textbf{Evaluation}} \\
  \multicolumn{1}{c|}{} &
  \multicolumn{1}{c|}{} &
  Entity &
  Event &
  Topic &
  Thread &
  Tree &
  Map &
  Case Study &
  User Study &
  \multicolumn{1}{c}{Ql/Qt} \\ \hline
\multicolumn{1}{l|}{\teaserLink{\textit{1}} ThemeRiver\cite{havre_themeriver_2002}} &
  News, Speeches &
   &
   &
  \checkmark &
  \checkmark &
   &
   &
  \checkmark &
  \checkmark &
  Ql \\
\rowcolor[HTML]{EFEFEF} 
\multicolumn{1}{l|}{\cellcolor[HTML]{EFEFEF}\teaserLink{\textit{2}} TextFlow\cite{cui_textflow_2011}} &
  Publications, News &
   &
   &
  \checkmark &
  \checkmark &
   &
   &
  \checkmark &
   &
  Ql \\
\multicolumn{1}{l|}{\teaserLink{\textit{3}} CloudLines\cite{krstajic_cloudlines_2011}} &
  News &
   &
  \checkmark &
   &
  \checkmark &
   &
   &
  \checkmark &
   &
  Ql \\
\rowcolor[HTML]{EFEFEF} 
\multicolumn{1}{l|}{\cellcolor[HTML]{EFEFEF}\teaserLink{\textit{4}} EventRiver\cite{luo_eventriver_2012}} &
  News &
   &
  \checkmark &
  \checkmark &
  \checkmark &
   &
   &
  \checkmark &
  \checkmark &
  Ql/Qt \\
\multicolumn{1}{l|}{\teaserLink{\textit{5}} Metro Map\cite{shahaf_metro_2012, shahaf_information_2013}} &
  News &
   &
  \checkmark &
  \checkmark &
   &
   &
  \checkmark &
   &
  \checkmark &
  Ql \\
\rowcolor[HTML]{EFEFEF} 
\multicolumn{1}{l|}{\cellcolor[HTML]{EFEFEF}\teaserLink{\textit{6}} StoryFlow\cite{liu_storyflow_2013}} &
  Movies, Twitter &
  \checkmark &
  \checkmark &
   &
  \checkmark &
   &
   &
  \checkmark &
   &
  Ql \\
\multicolumn{1}{l|}{\teaserLink{\textit{7}} EvoRiver\cite{sun_evoriver_2014}} &
  News &
   &
   &
  \checkmark &
  \checkmark &
   &
   &
  \checkmark &
  \checkmark &
  Ql \\
\rowcolor[HTML]{EFEFEF} 
\multicolumn{1}{l|}{\cellcolor[HTML]{EFEFEF}\teaserLink{\textit{8}} ThemeDelta\cite{gad_themedelta_2015}} &
  News &
   &
   &
  \checkmark &
  \checkmark &
   &
   &
  \checkmark &
  \checkmark &
  Ql \\
\multicolumn{1}{l|}{\teaserLink{\textit{9}} TimeSets\cite{nguyen_timesets_2016}} &
  Publication, News &
   &
  \checkmark &
  \checkmark &
  \checkmark &
   &
   &
  \checkmark &
   &
  Ql/Qt \\
\rowcolor[HTML]{EFEFEF} 
\multicolumn{1}{l|}{\cellcolor[HTML]{EFEFEF}\teaserLink{\textit{10}} StoryCurve\cite{kim_visualizing_2018}} &
  Movies &
  \checkmark &
  \checkmark &
   &
  \checkmark &
   &
   &
  \checkmark &
  \checkmark &
  Ql/Qt \\
\multicolumn{1}{l|}{\teaserLink{\textit{11}} StoryGraph\cite{ansah_graph_2019}} &
  Twitter &
   &
  \checkmark &
  \checkmark &
   &
  \checkmark &
   &
   &
   &
  Ql \\
\rowcolor[HTML]{EFEFEF} 
\multicolumn{1}{l|}{\cellcolor[HTML]{EFEFEF}\teaserLink{\textit{12}} StoryForest\cite{liu_story_2020}} &
  News, Blogs &
   &
  \checkmark &
   &
   &
  \checkmark &
   &
  \checkmark &
   &
  Ql/Qt \\
\multicolumn{1}{l|}{\teaserLink{\textit{13}} Narrative Map\cite{keith_norambuena_narrative_2021,keith_norambuena_design_2022,keith_norambuena_mixed_2023}} &
  News &
   &
  \checkmark &
   &
   &
   &
  \checkmark &
  \checkmark &
   &
  Qt \\
\rowcolor[HTML]{EFEFEF} 
\multicolumn{1}{l|}{\cellcolor[HTML]{EFEFEF}\teaserLink{\textit{14}} Narrative Graph\cite{yan_narrative_2023}} &
  News &
   &
  \checkmark &
  \checkmark &
   &
   &
  \checkmark &
  \checkmark &
   &
  Ql/Qt \\
\end{tabular}%
}
% \vskip 5pt 
\end{table*}

%% file: inputs/02_taxonomy/02_elements.tex
\begin{figure}[b]
    \centering
    \includegraphics[width=\linewidth]{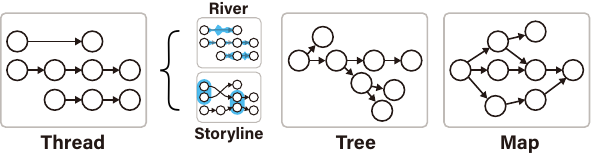}
    \caption{Representative narrative structures identified in our literature survey. They have in common that one spatial dimension indicates time (here, left-to-right). Different structures afford different ways to show relationships between announcements. While all use linear elements to connect pairs of announcements, the non-time dimension (here vertical) can show secondary relationships (such as shared events in storyline views) or strict thematic separation (such as thread views). Note that River is a subtype of Thread that additionally encodes time-dependent quantity (typically term frequency).}
    \label{fig:threadtreemap}
\end{figure}
\subsection{How is Time-Dependent Text Mapped to Visual Structures?}
In the context of news media and information analysis, we focus on a specific form of narrative: the structure of time-dependent text, that is, the interconnected progression of news stories across time and themes. Unlike traditional narratology, which studies storytelling within single documents (see \cite{santana2023survey} for a survey of single-document narrative extraction), we examine how multiple news articles connect and evolve. Building on definitions from a recent survey of news narrative extraction techniques \cite{norambuena_survey_2023}, we define \emph{narrative} as:
\textit{a temporal sequence of news announcements that are thematically connected through shared elements (entities, events, or topics).} We depart from the event-based focus of \cite{norambuena_survey_2023} by taking \emph{announcements} as the fundamental unit rather than events, which allows a single announcement to reference multiple events and the same event to be mentioned across articles.

To ensure consistency, we formalise the recurring elements and structures found in the literature into two components of our taxonomy: Story Context Elements (SCEs) and Narrative Data Topologies (NDTs).
\label{sec:narrative_taxonomy}
\subsubsection*{Story Context Elements (SCEs)}
The fundamental building blocks in time-dependent text visualisation are \textbf{announcements}.

\noindent\textbf{Announcement}: The smallest unit of text information (depending on analysis scope, e.g., sentence, document, or cluster). An announcement occurs at a specific time point or within a time window of arbitrary precision.

\noindent\textbf{Grouping attributes}: properties used to organise announcements into comparable sets.
\begin{itemize}
    \item \textit{Entity}: distinct individuals, objects, locations, or organisations with unique attributes.
    \item \textit{Event}: significant occurrences involving entities, including details such as who, what, where, and when.
    \item \textit{Topic}: overarching subjects that connect entities and events.
\end{itemize}

\noindent\textbf{Ordering attributes}: properties that determine progression.
\begin{itemize}
    \item \textit{Real time}: publication/event time in the source data.
    \item \textit{Narrative time}: presentation order in the viewed sequence.
\end{itemize}

\noindent\textbf{Track}: a visual band on the non-time axis used to group announcements by one organising attribute (for example, entity or topic).

In short, announcements are positioned by ordering attributes and grouped by selected content attributes. These SCEs appear frequently in different approaches. For example, TextFlow\cite{cui_textflow_2011} focuses on topic evolution, Story Forest\cite{liu_story_2020} emphasises event relationships, StoryFlow\cite{liu_storyflow_2013} tracks entity relationships, and Narrative Maps\cite{keith_norambuena_narrative_2021} combines multiple elements.

\subsubsection*{Narrative Data Topologies (NDTs)}
After defining elements, we define how they connect. Existing techniques mostly instantiate three NDT categories (\autoref{fig:threadtreemap}):
\begin{description}
    \item[Thread (Linear, $1 \rightarrow 1$)] Represents narratives without branches or merges, e.g., timelines\cite{brehmer_timelines_2017}. \textbf{Theme-River} and \textbf{Story-Line} are common thread subtypes with different non-time axis semantics.
    \item[Tree (Hierarchy, $1 \rightarrow N$)] Captures branching narratives and dependencies, as in Story Forest\cite{liu_story_2020}.
    \item[Map (Network, $N \leftrightarrow N$)] Depicts branching and merging narratives through spatial graph layouts, as in Information Cartography\cite{shahaf_information_2015}.
\end{description}

Hybrid systems can mix these categories, but the central distinction remains structural: whether a path stays single, branches only, or both branches and merges. This taxonomy by SCE and NDT is applied to representative time-dependent text visualisation systems in \autoref{tab:narrative_taxonomy}.

%% file: inputs/02_taxonomy/03_evaluation.tex
\subsection{Evaluating Time-Dependent Text Visualisation}
\label{sec:val_ntv}

Past evaluations of time-dependent text visualisation techniques typically use both qualitative and quantitative methods, but tend to focus on usability of specific visualisation idioms and data corpora with results that are difficult to generalise. Table~\ref{tab:narrative_taxonomy} summarises the evaluation approaches applied to various visual idioms.

\begin{figure*}
    \centering
    \includegraphics[width=\linewidth]{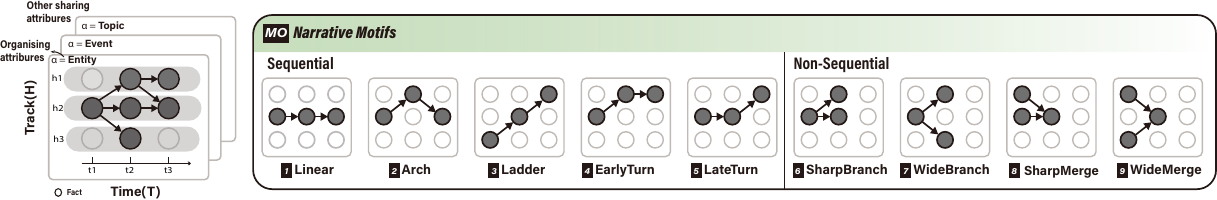}
    \caption{Left: An illustration of a Time-Track Narrative Graph Model. Right: The three node motifs for the Time-Track Narrative Graph Model include Sequential motifs (Linear, Arch, Ladder, Early Turn, Late Turn) and Non-Sequential motifs (Sharp Branch, Wide Branch, Sharp Merge, Wide Merge).}
    \label{fig:ttgraph}
\end{figure*}

ThemeRiver\cite{havre_themeriver_2000} used formative usability testing to assess user comprehension of the dynamic thematic frequency, collecting feedback via verbal protocols and questionnaires. TextFlow\cite{cui_textflow_2011} and StoryFlow\cite{liu_storyflow_2013} used semi-structured expert interviews to provide feedback on the effectiveness of the system in modelling, navigating, and analysing evolving topics. EventRiver\cite{luo_eventriver_2012} combined comparative experiments and user feedback with quantitative metrics to improve event detection and visualisation.
Metro Map\cite{shahaf_information_2015} and Narrative Maps\cite{keith_norambuena_narrative_2021} used user studies and quantitative metrics (cohesion, coherence, and coverage) to evaluate user understanding. EvoRiver\cite{sun_evoriver_2014} relied on user interactions and case studies to evaluate visualisation design and interactions.
Story Forest\cite{liu_story_2020} used comparative experiments with metrics such as coherence and readability, while StoryGraph\cite{ansah_graph_2019} employed quantitative methods using ROUGE scores.

As per the \emph{Evaluation} column in Table~\ref{tab:narrative_taxonomy}, most evaluations are by case study, with only a handful of user studies comparing the efficacy of specific visual idioms applied to news corpora data.

\subsection[]{\teaserLink{$T \rightarrow H \rightarrow G$}  Evaluating Human Understanding of Text}
\label{sec:eva_ntv}

The results of the studies above are limited to the tested visualisations and lack support in understanding how humans interpret narrative structure from text in general. Other studies explore how humans understand connectivity across text in a way that may give broader insight into time-dependent text visualisation design. Endert et al.\cite{endert_semantics_2012} studied 12 participants manually organising 200 documents with a visualisation tool, finding that they preferred clustering by higher-level concepts over keyword co-occurrence. Lee et al.\cite{lee_human_2017} highlighted the struggle of non-experts to interpret ranked word lists. Norambuena et al.\cite{norambuena_narrative_2021} studied 10 analysts manually constructing narrative maps from news articles, identifying different cognitive strategies to connect events in human-generated maps.

However, precise control of the text being visualised is required to evaluate text visualisation models more generally. Therefore, we need a more rigorous and controllable evaluation framework.

%% file: inputs/03_model/01_overview.tex
\section{\teaserLink{$TTNG$} Time-Track Narrative Graph Model}
\label{sec:tt_graph_model}

Building upon the elements identified in our literature review, we now formalise these concepts into a unified model (left side of Figure~\ref{fig:ttgraph}). Our Time-Track Narrative Graph (TTNG) provides a structured computational framework for understanding the fundamental structure of narratives in text, independent of any specific visualisation technique. Related abstractions exist in other text analysis scenarios; for example, Yousef and Janicke propose a directed graph model for text alignment visualisations\cite{yousef2020survey}. While their model operates at a very fine-grained (word-level) scale suited to alignment, our focus is a higher-level model for narrative structures in news.

Conceptually, we treat narrative visualisation as mapping relationships among evolving events, entities, and topics into graphical form so that connections cohere as visual “tracks” interpretable via Gestalt principles. We analyse these mappings in light of our findings on the understandability of TTNG structures.

\subsection{Narrative Space}

TTNG defines a narrative space $\mathcal{N} = T \times H$, where $T$ is the sequence of time windows and $H$ is the set of tracks formed by grouping announcements via the organising function $\alpha$, defined below. A \textit{track} is a visual band on the non-time axis used to organise announcements by a selected grouping attribute.
\subsection{Graph Structure}

The narrative in TTNG is defined as a graph $ G = (V, E) $, where $ V $ is the set of nodes (each representing an Announcement corresponding to one or more underlying texts produced or ``published'' within a specific window), and $ E $ is the set of edges representing select relationships between pairs of Announcements. These connections can form within a track or cross tracks, depending on which attributes are most significant to the narrative flow.

To make the construction explicit, we define TTNG structure as $\mathcal{S} = (\alpha, \beta)$, where $\alpha$ is the organising rule that assigns nodes to tracks, and $\beta$ is the linking rule that determines whether an edge is present between candidate node pairs.

%% file: inputs/03_model/02_definition.tex
\subsubsection*{Nodes (V)} A node $ v_i \in V $ is defined as a tuple $v_i = (p_i, A_i)$, where:
\begin{itemize}
      \item $p_i$ is the ordering attribute of node $v_i$ (typically time $t_i \in T$, or an equivalent sequence index).
      \item $A_i$ is the set of grouping attributes associated with $v_i$, such as entities and topics.
\end{itemize}

\subsubsection*{Track Assignment (H)} Nodes are assigned to tracks by a rule we call $\alpha$ (the organising rule). Common choices are by entity (each track is an entity), by magnitude (each track is a prominent theme/topic), or by structure (tracks follow optimised story paths). $\alpha$ determines the position on the non-time axis; time is fixed on the other axis.
\paragraph{Track Priority} A node may match multiple tracks. In such cases, we resolve the conflict with a track-priority order and assign the node to the highest-priority matching track.
% For each node $v_i$, the attribute set at level $\alpha$ is $a_i^\alpha$. Nodes are assigned to tracks corresponding to their attributes: $H_i = \{ h_a \in H \mid a \in a_i^\alpha \}$. A node can belong to multiple tracks if it contains multiple relevant attributes.

\subsubsection*{Edges (E)} We draw an edge between two announcements when they share a chosen attribute or relation. This choice is controlled by $\beta$ (the linking rule), operationalised as a predicate over node pairs, $\beta: V \times V \rightarrow \{0,1\}$. Cross-track edges typically arise when $\beta$ uses a \emph{different} attribute than $\alpha$ (e.g., tracks by topic via $\alpha$, but edges by shared entity via $\beta$). If $\beta$ uses the same attribute as $\alpha$, links are typically within-track. In practice, Thread/Storyline systems (e.g., ThemeRiver and Storyline) often omit explicit links\cite{havre_themeriver_2000,tanahashi_design_2012}, whereas Map-like systems (e.g., Narrative Maps, Information Cartography) draw explicit edges to reveal structure\cite{keith_norambuena_narrative_2021,shahaf_information_2015}.

\paragraph{\boldmath $\alpha$ vs.~$\beta$ in one line each}
\textbf{$\alpha$:} places each node on a track (y-position) determined by entity, magnitude (topic prominence, keyword frequency), or structure (optimised paths). \textbf{$\beta$:} adds explicit links between nodes. Use it when the visualisation needs visible cross-node connectivity; otherwise rely on track continuity.

\subsubsection*{Visual and Spatial Constraints}
From the designs of the surveyed visualisations, we infer the following constraints on TTNG structure, i.e., limitations on valid tracks and edges under $\alpha$ and $\beta$.
\begin{description}
      \item[Temporal Constraint (Hard)] Edges follow temporal order ($t_j > t_i$), ensuring acyclic graphs and strictly forward connections, and disallowing links at the same time point ($i = j$).
      \item[Track Exclusivity (Soft)] Announcements are assigned to only one track. Almost all surveyed examples follow this constraint, presumably to keep nodes within convex regions with simple colour mappings. One exception is \textit{Metro Maps}\cite{shahaf_metro_2013}, where nodes can lie on the border between track regions.
      \item[Cross-Track Adjacency Preference (Soft)] Prefer links between adjacent tracks and use long-distance cross-track links only when the narrative gain is clear, because long links increase layout and interpretation cost.
\end{description}

With these definitions and constraints, all of the surveyed narrative visualisation techniques can be modelled using TTNG. That is,
TTNG captures the inherent structure of narrative news text, regardless of how it might eventually be visualised (i.e.\ stylistic and metaphor concerns). By isolating the underlying narrative structure from all but the most basic spatial visualisation concerns, TTNG allows us to study fundamental patterns in how narratives develop and connect.

%% file: inputs/03_model/03_motifs.tex
\section[]{\teaserLink{$MO$} Narrative Motifs}
\label{sec:narrative_motif_identification}

As we see from the examples surveyed in \autoref{sec:background}, instances of TTNG visualisations may grow arbitrarily large. In this section, we distil minimal structural components—\emph{motifs}—that capture how narratives progress within and between tracks and can be composed into larger visualisations.\footnote{We call these fundamental TTNG components \textit{motifs}, but we note that other definitions of motifs as interesting subgraph structures have been used for analytical purposes in a variety of fields, including text analysis\cite{arnold2018advanced}.}

\paragraph{Why motifs?}
Motifs serve as small, composable units that capture the basic ways a narrative can evolve across time ($T$) and tracks ($H$) in TTNG. Using compact units offers two benefits: (i) representation—larger TTNGs can be expressed as compositions of a small set of recurring patterns; and (ii) evaluation—readers' ability to recognise these basic patterns provides a lower‑bound on the understandability of more complex structures built from them. Our user task in \autoref{sec:experimental_design} operationalises this idea by asking participants to select the motif whose structure best matches a short set of announcements.

\begin{figure*}
    \centering
    \includegraphics[width=\textwidth,trim=0mm 0mm 0mm 0mm]{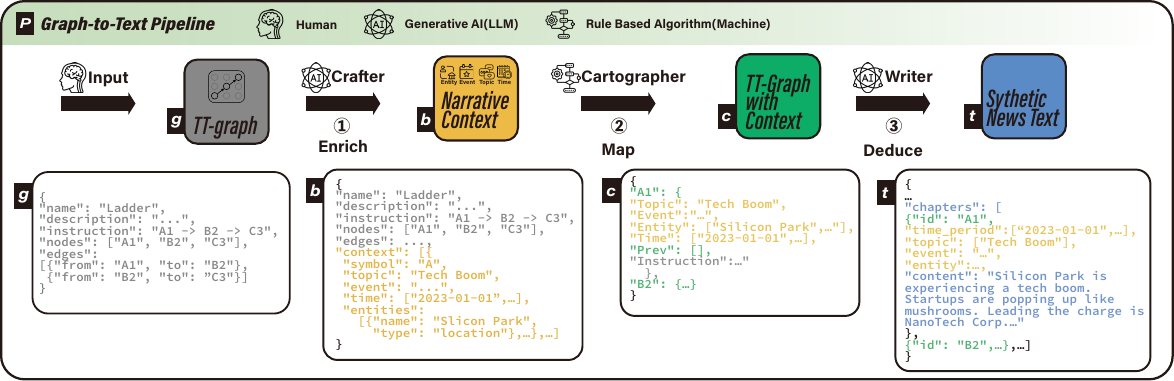}
    \caption{The \textbf{Graph-to-Text Pipeline}. This diagram illustrates the process of converting a TTNG into synthetic news text through three main stages: (1) Enriching the graph with narrative context using the Crafter; (2) Mapping narrative context elements with the Cartographer; (3) Generating the final synthetic news text with the Writer. The process enforces structural patterns through shared attributes while allowing natural thematic connections to emerge. For additional examples of narrative context, prompt templates, and the map algorithm, refer to the supplementary materials \ref{appendix:llm-graph-text-generation}.}
    \label{fig:graph_to_text_ct}
\end{figure*}

\subsection{Motif Constraints}

For simplicity and tractability, we impose the following constraints (in addition to the general visual and spatial constraints defined earlier for TTNG). These choices make the pattern space finite and enumerable for study while preserving the core phenomena of within‑track continuity and cross‑track transitions:

\begin{description}
    \item[3 nodes] are sufficient to express the canonical cases of continuation, turn, branch, and merge in our setting, while keeping cognitive and experimental complexity manageable.
    \item[3x3 grid] nodes within each motif lie in a 3x3 grid so that time is discretised into three positions and at most three tracks are shown. This yields a closed set of distinct patterns for three nodes and enables exact enumeration and balanced stimulus construction.
    \item[Time translation invariance] motifs that have the same structure but are offset in the time dimension of the 3x3 grid are considered the same.
    \item[Track order invariance] two motifs are considered the same if they can be matched via track reordering.
    \item[Track Linearity] Tracks are presented linearly within fixed bands in the non-time (vertical by default) dimension.
\end{description}

Subject to these constraints on motifs, we can enumerate precisely nine basic patterns of narrative flow in a 3x3 time $T$ and track $H$ space. These motifs constitute the canonical building blocks we evaluate for understandability (\autoref{sec:results}). If readers struggle with these basic patterns, we expect similar difficulties with more complex structures that compose them.

Motifs can be composed (e.g., by joining two motifs at a common Announcement) to construct larger TTNGs. Full composition rules are out of scope; here we focus on the understandability of basic motifs. Relaxing track exclusivity enables set overlays and covers many surveyed examples.

\paragraph{Algorithmic Enumeration (Appendix)}
We enumerate motifs under TTNG constraints using a canonical occupancy-matrix formulation (exact node count, minimum temporal span, and left-packed canonical form), with track-permutation equivalence to avoid duplicates. The full derivation, including recursive construction from $f(m,n,X)$ to $f(m,n,X-1)$ and justification for the 3-node base case, is provided in \autoref{appendix:narrative_motif_identification}.

We subdivide the nine motifs into \textit{sequential} and \textit{non-sequential} categories based on track transitions, see right side of Fig.\ \ref{fig:ttgraph}.

\subsection{Sequential Motifs}

Sequential motifs show ways narratives can progress through time:

\begin{description}
    \item[Linear] \raisebox{-0.2\height}{\includegraphics[height=1.2em]{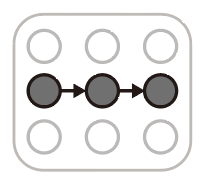}} Announcements within single track, showing that the organising attribute remains most significant $(h_i, t_j) \rightarrow (h_i, t_{j+1}) \rightarrow (h_i, t_{j+2})$.

    \item[Arch] \raisebox{-0.2\height}{\includegraphics[height=1.2em]{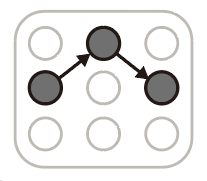}} Narrative briefly transitions to another track when a shared attribute becomes significant, then returns when the organising attribute regains significance $(h_i, t_j) \rightarrow (h_{i+1}, t_{j+1}) \rightarrow (h_i, t_{j+2})$.

    \item[Ladder] \raisebox{-0.2\height}{\includegraphics[height=1.2em]{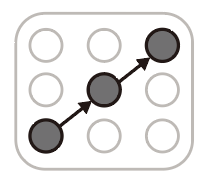}} Narrative progressively moves across tracks as different attributes gain significance $(h_i, t_j) \rightarrow (h_{i+1}, t_{j+1}) \rightarrow (h_{i+2}, t_{j+2})$.

    \item[EarlyTurn] \raisebox{-0.2\height}{\includegraphics[height=1.2em]{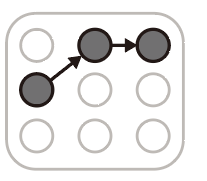}} A shared attribute quickly becomes significant, causing an early track transition that persists\\
          $(h_i, t_j) \rightarrow (h_{i+1}, t_{j+1}) \rightarrow (h_{i+1}, t_{j+2})$.

    \item[LateTurn] \raisebox{-0.2\height}{\includegraphics[height=1.2em]{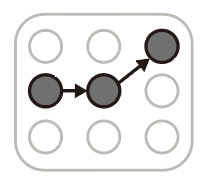}} The organising attribute remains dominant before a late transition when a shared attribute becomes significant\\
          $(h_i, t_j) \rightarrow (h_i, t_{j+1}) \rightarrow (h_{i+1}, t_{j+2})$.
\end{description}

\subsection{Non-Sequential Motifs}

Non-Sequential motifs show cases where multiple tracks become relevant simultaneously:

\begin{description}
    \item[SharpBranch] \raisebox{-0.2\height}{\includegraphics[height=1.2em]{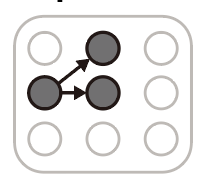}} An Announcement connects to two adjacent tracks because it has multiple significant attributes\\
          $(h_i, t_j) \rightarrow \{(h_i, t_{j+1}), (h_{i+1}, t_{j+1})\}$.

    \item[WideBranch] \raisebox{-0.2\height}{\includegraphics[height=1.2em]{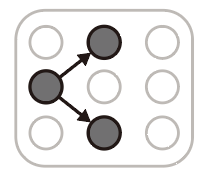}} An Announcement connects to distant tracks due to strongly contrasting attributes $(h_i, t_j) \rightarrow \{(h_{i+1}, t_{j+1}), (h_{i+2}, t_{j+1})\}$.

    \item[SharpMerge] \raisebox{-0.2\height}{\includegraphics[height=1.2em]{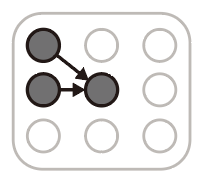}} Announcements from adjacent tracks converge when they share a significant attribute $\{(h_{i+1}, t_j), (h_{i}, t_j)\} \rightarrow (h_i, t_{j+1})$.

    \item[WideMerge] \raisebox{-0.2\height}{\includegraphics[height=1.2em]{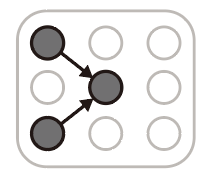}} Announcements from distant tracks converge when they develop a strong shared attribute $\{(h_i, t_j), (h_{i+2}, t_j)\} \rightarrow (h_{i+1}, t_{j+1})$.
\end{description}

These motifs serve as building blocks for understanding how readers follow narratives as they move within and between tracks. The next step is to use these motifs to generate text narratives that explore how readers respond to different transition patterns.

%% file: inputs/04_pipeline/01_overview.tex
\section[]{\teaserLink{$ G \longrightarrow L \longrightarrow T $} Graph-to-Text Pipeline}
\label{sec:graph_to_text}

The next challenge towards human-centred evaluation of time-dependent text visualisation is the absence of datasets with a known (predefined) narrative structure, which serves as ground truth for evaluating narrative extraction algorithms. To address this challenge, we propose a Graph-to-Text pipeline designed to produce controlled fictional news datasets, leveraging Large Language Models (LLMs) to accelerate study and analysis.

Control ensures that the generated texts align with the structural requirements of the TTNG model. For example, to evaluate perceptions of linear visual idioms, we might need a collection of news stories that unfold linearly within a coherent narrative.

A fictional dataset helps avoid confounds from prior knowledge: real-world news carries readers’ backgrounds, habits, and preconceived narratives that may skew perception analysis\cite{zhao_evaluating_2022}. Most existing fictional news datasets focus on clustering and summarisation \cite{levi_compres_2020,sultana_narrative_2022}, offering little in the way of clear narrative structure. Creating new datasets with complex structures is also resource-intensive, requiring expert annotation among other investments.

%% file: inputs/04_pipeline/02_implementation.tex
\subsection[]{\teaserLink{$P$} Pipeline Components}

Our Graph-to-Text pipeline (Fig.~\ref{fig:graph_to_text_ct}) converts a TTNG into synthetic news text through three main stages, each leveraging different computational approaches. As annotated in the figure banner, the Crafter and Writer use generative AI (LLM), while the Cartographer is rule-based.

\paragraph{Input: TTNG} The pipeline begins with a TTNG specification in JSON format, as shown in Fig.~\ref{fig:graph_to_text_ct} (panel \textit{g}). This defines the narrative structure with nodes ("nodes": ["A1", "A2", "A3"]) and their connections ("edges": [["A1", "A2"], ["A2", "A3"]]).

\paragraph{1. Crafter (Enrich)}
The Crafter enriches the basic graph structure with narrative context using GPT-4. As shown in Fig.~\ref{fig:graph_to_text_ct} (panel \textit{b}), it:
\begin{itemize}
    \item Generates a narrative space with entities, events, and topics (highlighted in yellow)
    \item Defines the thematic domain (e.g., "Tech Boom")
    \item Creates initial attribute mappings (e.g., "Silicon Park" location setting)
\end{itemize}

\paragraph{2. Cartographer (Map)}
The Cartographer (Fig.~\ref{fig:graph_to_text_ct}, panel \textit{c}) maps the enriched narrative elements into a structured format through:
\begin{itemize}
    \item \textit{Time Assignment}: Maps events to specific timestamps
    \item \textit{Attribute Mapping}: Associates nodes with specific entities, events, and topics
    \item \textit{Connection Planning}: Defines how nodes should connect through shared attributes
    \item \textit{Structure Validation}: Ensures the mapping adheres to TTNG constraints
\end{itemize}

\paragraph{3. Writer (Deduce)}
The Writer (Fig.~\ref{fig:graph_to_text_ct}, panel \textit{t}) generates the final news text by:
\begin{itemize}
    \item Following the original graph structure
    \item Maintaining narrative coherence through shared attributes
    \item Reading naturally while preserving intended narrative connections
\end{itemize}

\paragraph{Output} The final output is a collection of synthetic news articles (shown in blue in Fig.~\ref{fig:graph_to_text_ct}, panel \textit{t}) that:
\begin{itemize}
    \item Follow the specified narrative structure
    \item Maintain thematic coherence
    \item Read naturally while preserving intended connections
\end{itemize}

This pipeline ensures structural patterns are preserved while allowing natural thematic connections to emerge. For more details about the Graph-to-Text pipeline, please refer to our source code on GitHub repository at \url{https://github.com/SonghaiFan/gpt_storytale}.

%% file: inputs/04_pipeline/03_validation.tex
\subsection{Basis for Claims on GPT Model's Understanding}

Previous work in narrative analysis has evolved from simple text similarity metrics to sophisticated event-based approaches. Early works like TextFlow \cite{cui_textflow_2011} extended topic modelling techniques to capture temporal thematic changes, while Story Forest \cite{liu_story_2020} introduced semi-supervised document clustering using keyword co-occurrence graphs. Information Cartography \cite{shahaf_information_2015} and Narrative Maps \cite{norambuena_narrative_2021} employ optimisation techniques with TF-IDF and embedding-based classifiers. Building on these approaches, we validate our pipeline's output using three complementary metrics that align with this progression:

\begin{description}
\item[Jaccard Similarity] Quantifies entity-level overlap between Facts
\item[TF-IDF Cosine Similarity] Measures term-level coherence
\item[BERT Similarity] Captures semantic-level relationships
\end{description}

The narrative structure in TTNG is reflected by how consecutive Announcements follow the intended track shifts. Ideally, in a LateTurn TTNG with Topic-focused Track, Announcements 1--2 should show high BERT similarity scores but drop between 2--3, reflecting the late topic shift. On the other hand, in an Entity-LateTurn, Announcements 1--2 should maintain high Jaccard entity similarity but decrease significantly for 2--3 as the primary entities change.

The results shown in Fig.~\ref{fig:machine_understanding_box} confirm these expected patterns across all metrics, validating our pipeline's ability to generate narratives with controlled structures and meaningful track transitions. This pairwise similarity analysis provides a straightforward and quantifiable way to assess the quality of generated datasets and verify that the intended narrative structures are properly reflected in the text.

%% file: inputs/05_study/01_design.tex
\section{User Study Design}
\label{sec:experimental_design}

Our study explores how people interpret narrative structure in news text as a graph. Participants were asked to select the narrative motif (3-node TTNG pattern as defined in Section~\ref{sec:narrative_motif_identification}) that best matched the presented news text (announcements), which was generated from predefined motifs as ground truth.

\subsection{Overview of the Synthetic Dataset}

The narrative structure used in this study is a 3-node TTNG, where each node is an announcement (a short news abstract).

To counteract randomness caused by generative AI (GPT-4), the user study was designed as three comparative sets. Each set contains announcements illustrating all \textbf{nine} predefined \textit{narrative motifs} (Linear, Arch, Ladder, etc.). Additionally, \textbf{one} common control story (with a predefined Ladder motif) is included in each set for cross-set comparison.

We generate $9 \times 3 + 1 = 28$ distinct news stories, each consisting of three Announcements (following our 3-node TTNG model), resulting in a total of $27 \times 3 + 1 \times 3 = 84$ Announcements.

%% file: inputs/05_study/02_validation.tex
\subsection{Validation of the Synthetic Dataset}
\label{sec:datasetvalidation}

To ensure the quality of our synthetic dataset for the user study, we analysed the similarity distributions between Announcements using  previously established metrics (Jaccard, TF-IDF, and BERT). 

\paragraph{Statistical Significance Testing}
Two-sample t-tests performed on each metric yielded p-values < 0.001, confirming statistically significant differences between same-track and different-track Announcement pairs. The clear separation in similarity scores validates that our synthetic dataset maintains the intended track distinctions, at least from an algorithmic metric perspective (see Fig \ref{fig:machine_understanding_box}):

\begin{figure}
\centering
\includegraphics[width=1\linewidth,trim=0mm 0mm 0mm 0mm]{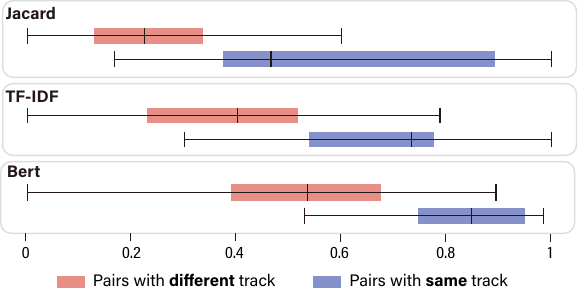}
\caption{Distribution of similarity scores for Announcement pairs in our synthetic dataset. Box extents give 25th to 75th percentiles, vertical line gives median, whilst extent of horixontal lines indicate min--max. Red boxes show similarity scores for Announcements from different tracks (expected to be lower), while blue boxes show scores for Announcements in the same track (expected to be higher). Across all three metrics (Jaccard, TF-IDF, and BERT), there is clear separation between same-track and different-track pairs, with same-track pairs consistently showing higher similarity scores compared to different-track pairs.}
\label{fig:machine_understanding_box}
\end{figure}

\begin{description}
    \item[Jaccard] Same-track pairs show 0.4-0.9 similarity, while different-track pairs show 0.1-0.4
    \item[TF-IDF] Same-track pairs cluster around 0.5-0.8, different-track pairs around 0.2-0.45
    \item[BERT] Shows the clearest separation with same-track pairs at 0.75-0.95 and different-track pairs at 0.4-0.7
\end{description}

%% file: inputs/05_study/03_procedure.tex
\subsection{Procedure}

Our study recruited native English speakers via Prolific, a crowd-sourcing research platform that connects researchers with participants for studies and surveys. Gender distribution was balanced ($57$\% female, $43$\% male) among $30$ participants, aged $20$ to $57$. Educational backgrounds included undergraduate degrees ($37$\%), high school diplomas/A-levels ($30$\%), and graduate degrees ($20$\%). Reading habits were self-reported on a Likert scale (1 = rarely reads news, 5 = reads daily), with participants spanning levels 1 to 4.

Participants were first introduced to the TTNG Model through an instructional page. Following this, a hands-on training session was conducted in which participants read news Announcements, displayed on the right side of the webpage in plain text format (content with timestamp), and matched them to one of the nine motifs listed on the left.

During the training phase, participants were required to correctly identify motifs before proceeding to the next task. Each training task corresponded to a distinct story characterised by a predefined motif. After participants became familiar with the task, they moved to the main task. We encouraged participants to focus on making decisions based on their understanding, rather than worrying about being right or wrong, provided they gave clear reasoning.

In the task phase, participants selected the closest motif from 9 options based on their understanding of the text, \textbf{without} hints such as titles or keywords.
For a total of 10 tasks, story order was \textbf{randomised} to reduce order bias. Participants recorded their reasoning and self-assessed confidence for each selection.

\subsection{Data Collection and Analysis}

Data was anonymously recorded through the Google Analytics cloud service. The collection of data for analysis comprised both quantitative data, such as selection and time allocated per task, and qualitative data from the participants' reasoning for their selection. Task engagement times ranged from $12.39$ to $95.96$ seconds, averaging $53.04$ seconds.

%% file: inputs/06_results/01_overview.tex
\begin{figure}
      \centering
      \includegraphics[width=\linewidth,trim=0mm 0mm 0mm 0mm]{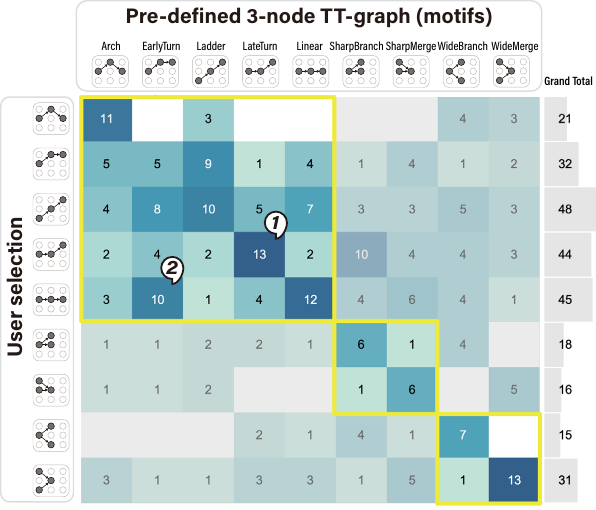}
      \caption{Confusion matrix comparing participant-selected motifs (rows) against predefined motifs (columns). Diagonal cells indicate correct identifications. The Grand Total column shows total selections per motif. Each column sums to 30, controlled by the graph-to-text pipeline input. Numbered talk bubbles correspond to detailed analysis in \ref{sec:case}.}
      \label{fig:confmat}
\end{figure}
\section{User Study Results}
\label{sec:results}

Our study reveals significant challenges in how readers interpret non-sequential narrative motifs. On average, participants correctly identified $3.1$ of the 10 motifs used to generate stories, with significant variance observed among participants. The best-performing individual accurately identified $7$ out of $10$ motifs, while the majority of participants ($22$ out of $30$) recognised $3$ or fewer motifs. This indicates a broad difficulty in recovering predefined narrative patterns, even in controlled synthetic narratives. In TTNG terms, readers tended to follow implicit continuity along tracks (Gestalt continuity) and struggled to perceive explicit cross-track edges. This suggests minimising unnecessary cross-track edges and, where possible, aligning $\alpha$ and $\beta$ at the same SCE level or using the same attribute selection criteria. In the following sections, we discuss the
motifs that were most and least accurately identified and some sources of confusion
identified from the qualitative data collected.

% \begin{figure*}[b!]
%     \centering
%     \begin{minipage}{0.49\textwidth}
%         \centering
%         \includegraphics[width=\linewidth]{figs/case1.pdf} % Example image
%         \caption{Confusion case study for LateTurn}
%         \label{fig:confusion_case1}
%     \end{minipage}\hfill
%     \begin{minipage}{0.5\textwidth}
%         \centering
%         \includegraphics[width=\linewidth]{figs/case2.pdf} % Example image
%         \caption{Confusion case study for EarlyTurn}
%         \label{fig:confusion_case2}
%     \end{minipage}
% \end{figure*}

%% file: inputs/06_results/02_quantitative.tex
\subsection{Quantitative Results}
\label{sec:insight}

\noindent\textbf{Track Transition Recognition Patterns.}
Analysis of the confusion matrix (Figure~\ref{fig:confmat}) reveals several patterns in how readers recognised narrative track transitions in our study.
Looking first at the ``Grand Total'' column, we see a bias in participant selections towards sequential motifs. With the exception of the final row/column (WideMerge), participants were more likely to select sequential motifs than non-sequential motifs in this task. The \textit{Ladder} motif received $48$ selections, followed by \textit{Linear} with $45$ selections.

\noindent\textbf{Recognition Accuracy.}
The timing and clarity of track transitions appeared to impact recognition in our sample:
\begin{itemize}
      \item \textit{LateTurn} and \textit{WideMerge} showed the highest accuracy (both 43\%), which may indicate that clearer, well-established track transitions are easier to recognise
      \item \textit{Linear} progression was also recognised relatively well (40\% accuracy), consistent with readers' familiarity with linear narratives
      \item Sequential motifs showed notable inter-confusion, particularly among \textit{EarlyTurn}, \textit{Linear}, and \textit{Ladder}
      \item In our dataset, temporal aspects were more consistently identified than track changes (59.2\% of selections matched temporal characteristics)
\end{itemize}

\begin{figure}
      \centering
      \includegraphics[width=\linewidth]{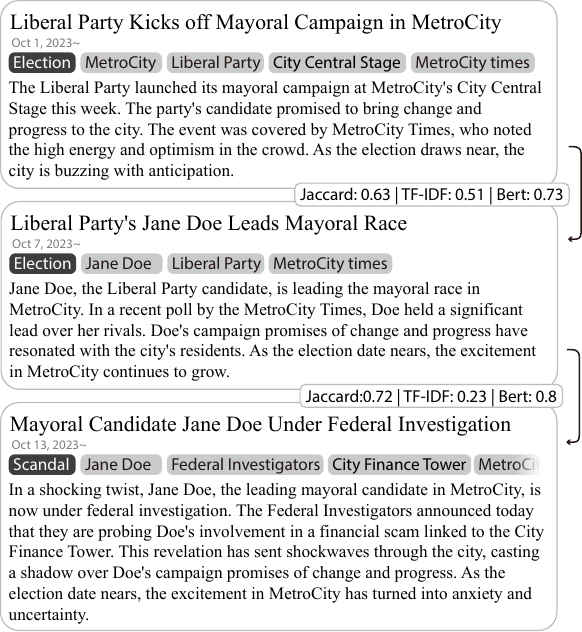}
      \caption{Confusion case study for LateTurn}
      \label{fig:confusion_case1}
\end{figure}
\begin{figure}
      \centering
      \includegraphics[width=\linewidth]{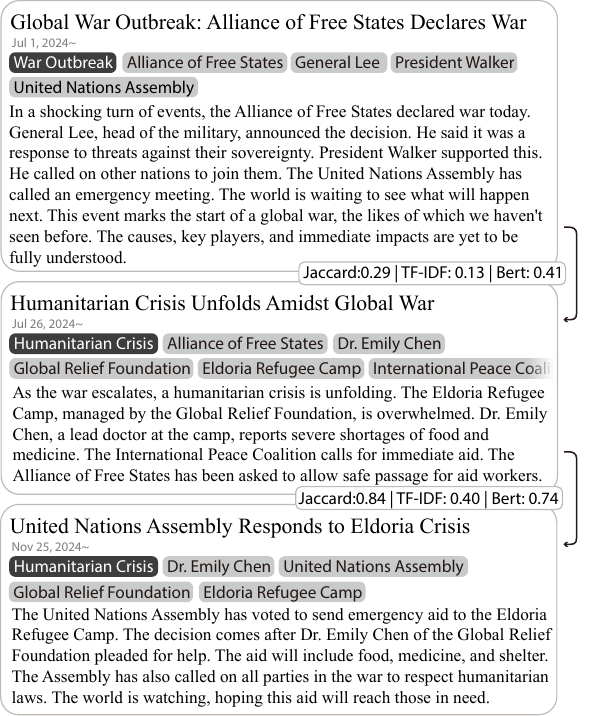}
      \caption{Confusion case study for EarlyTurn}
      \label{fig:confusion_case2}
\end{figure}

\subsection{Case Studies}
\label{sec:case}

We examine two contrasting cases from our confusion matrix (Figure~\ref{fig:confmat}):

\paragraph{Case A: \textit{LateTurn} recognition}
One of the more accurately identified motifs in set A (7/10 correct selections) showed a clear transition from Liberal party elections to financial fraud. Several participants described this as a ``jump'':

\begin{itemize}
      \item ``\textit{The twist shows a jump from optimism for Jane to the investigation}'' --- Participant A's reasoning focused on \emph{Entity Level Narrative}.
      \item ``\textit{The first two are regarding the candidates' progress in the mayoral election. The final story jumps to a federal investigation, no longer super focused on the election progress.}'' --- Participant B's reasoning focused on \emph{Event Level Narrative}.
\end{itemize}

\paragraph{Case B: \textit{EarlyTurn} misidentification}
This frequently misidentified motif showed confusion with \textit{Linear} (10 times) and \textit{Ladder} (8 times). While our metrics indicate a track shift, participants often perceived the progression as natural:

\begin{itemize}
      \item ``\textit{I believe it is linear because it goes from the declaration of war to humanitarian crisis, and to Aid being sent for humanitarian crisis. This seems like a linear progression of events following a declaration of war.}'' --- Participant C's reasoning led to a choice of Linear from Event level.

      \item ``\textit{Nonetheless I'm only moderately confident because the war concept does seem like lots is going on, even if it doesn't feel significant enough to impact the y changing}'' --- Participant D's reasoning led to a choice of Linear from Topic level.
\end{itemize}

\noindent We observe three tentative patterns about track transition recognition in this study:
\begin{itemize}
      \item Late transitions were more often recognised than early ones in our task, possibly because the initial track had been more clearly established
      \item When transitions were subtle, readers tended to default to simpler sequential patterns (e.g., \textit{Linear}, \textit{Ladder})
      \item Clear breaks in narrative focus (e.g., \textit{LateTurn}, \textit{WideMerge}) were more readily identified than gradual transitions
\end{itemize}
Overall, within this study's scope, recognition tended to be higher when transitions followed an established track (a track with prior narrative context from earlier nodes); early transitions were more often missed.

%% file: inputs/06_results/03_qualitative.tex
\begin{figure}
      \centering
      \includegraphics[width=\linewidth,trim=0mm 0mm 0mm 0mm]{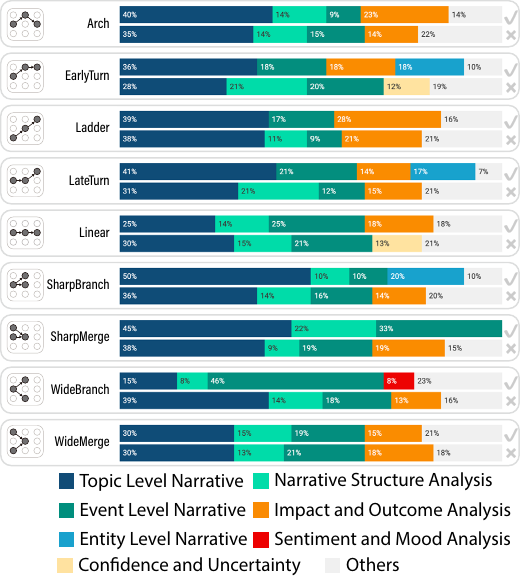}
      \caption{Distribution of reasoning categories for narrative selections. Top bars show distribution when participants correctly identified motifs, Bottom bars show distribution for incorrect identifications.}
      \label{fig:cat_bar}
\end{figure}

\subsection{Qualitative Analysis: Track Recognition Strategies}
\label{sec:ql_result}

Analysis of participant comments revealed several approaches to recognising track transitions, with a notable disparity between human and algorithmic strategies in this dataset.
\begin{description}
      \item[Track Recognition Hierarchy]
            As shown in Figure~\ref{fig:cat_bar}, participants often relied on higher-level SCEs for track identification in our study:
            \begin{itemize}
                  \item \textbf{Topic Level Analysis}: Identifying transitions based on topic changes (e.g., ``\textit{election to controversy...}'').
                  \item \textbf{Entity Level Analysis}: Following specific entities and their development (e.g., ``\textit{from Johnson's mention of his plans for improvement...}'').
                  \item \textbf{Event Level Analysis}: Tracking chronological progression of events (e.g., ``\textit{start of rebel group, then continuing fight with Davis's forces...}'').
                  \item \textbf{Structure Analysis}: Examining narrative patterns (e.g., ``\textit{since it follows a very linear progression...}'').
                  \item \textbf{Causality Analysis}: Identifying cause-effect relationships (e.g., ``\textit{the results of the previous statement directly leads into the next one...}'').
                  \item \textbf{Sentiment Analysis}: Noting sentiment progression (e.g., ``\textit{it went positive, negative, shifting towards positive...}'').
            \end{itemize}

      \item[Human vs Algorithmic Track Recognition]
            We observed a contrast between human and model approaches to track recognition. While our TTNG model relies heavily on entity-level connections for determining track transitions, participants gravitated towards higher-level narrative elements. This points to a potential gap between computational and cognitive approaches in this setting.

      \item[Causal Relationships in Track Recognition]
            Impact and outcome analysis was a recurrent theme for some participants when interpreting cross-track narratives. This focus on causality, particularly evident in event-level analysis, suggests that readers often seek cause-effect relationships when determining track transitions --- a factor not explicitly modelled in the current TTNG framework.

      \item[Extended SCE Dimensions]
            A minority of responses (approximately 8--12\%) referenced sentiment and mood, indicating an additional dimension of narrative understanding not currently captured in our SCE framework. This may serve as an additional signal for track transitions in human interpretation.
\end{description}
\noindent In this study, many readers relied more on higher-level narrative elements and causal relationships. These observations suggest potential directions for expanding the TTNG model to better align with human narrative comprehension patterns.

%% file: inputs/06_results/04_suggestion.tex
\subsection{Suggested Design Considerations Based on Insights}
\label{sec:suggestions}

Drawing on our quantitative and qualitative findings, we offer the following considerations for time-dependent text visualisation in the news domain. These are grounded in this study's tasks and data; applicability may vary by context and audience.

\begin{description}
      \item[Track Organisation and Transitions]
            In our study, readers often relied on higher-level SCEs (topic and event) for track recognition, while the model better detected entity-level connections. Consider:
            (1) Multi-level visualisation that shows both high-level topics and underlying entity connections;
            (2) Highlighting causal relationships between Announcements across tracks when available, as some readers used these to understand transitions;
            (3) Optionally incorporating sentiment/mood indicators as supplementary signals, where appropriate and reliable.

      \item[Track Transition Timing]
            In our study, late transitions (43\% accuracy) were better recognised than early transitions. Consider:
            (1) Allowing sufficient track history (more prior nodes) to establish track identity before major transitions;
            (2) Providing stronger visual cues for early transitions, which were more frequently missed in our task;
            (3) Using progressive disclosure to help readers anticipate upcoming changes when appropriate.

      \item[Sequential vs Non-Sequential Structures]
            In our task, readers showed a preference for sequential patterns (\textit{Ladder}: 48 selections; \textit{Linear}: 45 selections). Consider:
            (1) Favouring sequential flows when consistent with editorial goals;
            (2) Using visual emphasis to signal when non-sequential transitions are important.

      \item[Entity-Level Visualisation]
            While many readers focused on topic-level understanding, entity tracking appeared to help with more accurate motif identification. Consider:
            (1) Entity highlighting across tracks to show connections;
            (2) Interactive features to toggle between topic and entity views;
            (3) Visualising entity relationships that bridge different tracks, where data quality permits.

      \item[Temporal Aspect Enhancement]
            In our dataset, temporal aspects were identified more consistently than track changes (59.2\%). Consider:
            (1) Using clear temporal markers to anchor transitions;
            (2) Maintaining consistent temporal progression across multiple tracks;
            (3) Providing temporal context for transitions.

      \item[Attribute Significance Visualisation]
            To help explain why track transitions occur in a given context, consider:
            (1) Visually encoding the strength of organising attributes within tracks;
            (2) Showing when shared attributes become significant enough to warrant transitions;
            (3) Interactive tools to explore different attribute weightings.
\end{description}

These considerations aim to help bridge the observed gap between algorithmic track detection and human narrative comprehension patterns, within the limits of this study, while supporting both high-level understanding and detailed exploration of narrative structures.

%% file: inputs/07_discussion.tex
\section{Limitations and Future Work}
\label{sec:discussion}

Our work demonstrates how tracks and transitions can guide narrative visualisation design. While our model provides a framework for representing narrative progression through track transitions, several limitations need to be considered.

First, our \textit{Graph-to-Text Pipeline}, though innovative, is in its early development. The synthetic news datasets generated may lack the complexity of real-world articles, which affects the ecological validity of our findings. Refining the pipeline with advanced language models and diverse narrative motifs could address these issues. An iterative validation/re-generation step should be added to ensure synthetic datasets accurately reflect the intended temporal and track structures in the TT-graph, ensuring clear separation between 'same' and 'different' track compositions (refer to Figure~\ref{fig:machine_understanding_box}).

Second, our user study is exploratory and formative. We wanted to test whether participants would infer the structures used to generate the texts based on reading the text content alone, without specific details about the expected mappings to edges and tracks. Additional training in precisely how to detect edges (e.g., by enforcing understanding of the kind of similarity metrics used for our generated text validation \autoref{sec:datasetvalidation}) may have resulted in less of the confusion studied in \autoref{sec:results}. However, it was precisely this variation in interpretations that we were interested in and which informs our design recommendations \autoref{sec:suggestions}. Furthermore, we believe it is this kind of open-ended exploration of people's mental models of structure that will best inform future investigation into text visualisation methods that are either as universally understandable as possible, or adaptable to individual needs.

Third, our user study was conducted with a relatively small sample size (n=30), which may limit the generalisability of our findings. Future studies should aim to recruit a larger and more diverse pool of participants to enhance the robustness and universality of the results.

Fourth, as an initial step towards modelling narratives using graphs, we deliberately impose certain edge constraints to maintain simplicity. While real narratives often contain more complex structures---such as long-distance thematic connections, flashbacks, circular references, simultaneous events, and implicit relationships spanning multiple contexts---these limitations are intentional design choices that make the model more tractable for initial implementation and evaluation. Future work could explore relaxing these constraints to capture such complex narrative structures.
% Fourth, while task order was randomised, we did not employ counterbalancing (e.g., Latin square). Residual order, learning, or fatigue effects may therefore remain and could confound condition comparisons. Future work should adopt counterbalanced designs to better control order effects.

Finally, while our design suggestions offer promising directions, they need to be validated through more controlled user studies to establish their effectiveness. Using our framework and pipeline as a starting point would be helpful, but further empirical research is necessary to solidify these design recommendations, as well as to extend evaluation to larger TTNG visualisations that are the composition of motifs.

%% file: inputs/08_conclusion.tex
\section{Conclusion}
\label{sec:conclusion}

Our research presents the Time-Track Narrative Graph (TTNG) model as a framework for representing narrative structure and track transitions. Through our user study of synthetic narratives, we observed a gap between algorithmic track detection and human narrative comprehension patterns.

Our findings highlight several fundamental aspects of narrative understanding:
(1) In our study, many readers relied on high-level SCEs (topics/events) to identify tracks, while algorithms better detected entity-level connections;
(2) Track transitions tended to be more recognisable after establishing track identity;
(3) Some readers sought causal relationships and emotional progression across tracks—elements not explicitly encoded in our generation pipeline and generally underexplored in past work on narrative visualisation.

Taken together, while computational models effectively detect entity-level connections, many participants appeared to require additional signals to understand track transitions, including clear organising principles, causal relationships, and sentiment progression. This suggests that multi-level views connecting entity-level patterns with higher-level narrative elements may help bridge the gap in some contexts.

% This study points to a core design and representation challenge—how to describe and control knowledge structures at the Topic, Event, and Entity levels—and how effectively TTNG can incorporate and represent a \textit{Recognition Hierarchy} within its framework.

By addressing these challenges, we hope we can move towards developing narrative visualisations that better bridge the gap between computational detection and human comprehension, leading to more intuitive exploration of complex narratives.

%% file: additional.tex
\section{Generating Text from Graph using LLMs}
\label{appendix:llm-graph-text-generation}
This appendix demonstrates how a LLMs can be utilized to generate coherent and contextually relevant text based on a given graph structure. The process involves aligning story elements with the graph nodes and edges, and then using the LLMs to generate text for each node while maintaining narrative coherence.

\subsection{Graph Structure}
Consider the following simple Arch graph structure representing a narrative:
\begin{figure}[h]
  \centering
  \begin{tikzpicture}[
      node distance=1cm and 1cm, % X and Y distance
      every node/.style={thick, draw, circle},
      every path/.style={->, thick}
    ]
    % Use a single dark style for highlighted nodes.
    \tikzstyle{darknode} = [fill=black!55, draw=black!70, text=white]

    % Nodes
    \node[darknode] (A1) {A1};
    \node[darknode] (B2) [above right of=A1] {B2};
    \node[darknode] (A3) [below right of=B2] {A3};

    % Paths
    \draw (A1) -- (B2);
    \draw (B2) -- (A3);
  \end{tikzpicture}
  \caption{Example story graph structure}
  \label{fig:story-graph}
\end{figure}
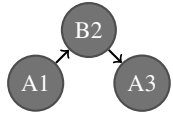

Each node in the graph represents a chapter or event in the story, and the edges represent the flow and dependencies between the nodes.

\subsection{Narrative Context}
\label{appendix:narrative-context}
\begin{lstlisting}[caption=JSON Context Example]
{
    "name": "Arch",
    "description": "The narrative begins with Theme A, then shifts to Theme B, and finally returns to Theme A.",
    "structure": "A1 -> B2 -> A3",
    "instructions": "Contains ONLY 3 nodes and 2 edges in total.",
    "nodes": ["A1", "B2", "A3"],
    "edges": [
      { "from": "A1", "to": "B2" },
      { "from": "B2", "to": "A3" }
    ],
    "context": [
      {
        "symbol": "A",
        "theme": "Economic Boom",
        "description": "The city experiences an unprecedented economic rise but faces a sudden ecological challenge before finding a sustainable economic solution.",
        "time": ["2023-01-01", "2023-12-31"],
        "entities": [
          {
            "name": "Mayor Lina",
            "type": "people"
          },
          {
            "name": "Downtown District",
            "type": "location"
          },
          {
            "name": "Economic Council",
            "type": "organisation"
          }
        ]
      },
      {
        "symbol": "B",
        "theme": "Ecological Crisis",
        "description": "Due to the rapid economic growth, the city faces an environmental crisis, leading to debates on the city's future.",
        "time": ["2023-06-01", "2023-09-01"],
        "entities": [
          {
            "name": "GreenLife",
            "type": "organisation"
          },
          {
            "name": "Riverstone Park",
            "type": "location"
          },
          {
            "name": "Dr. Emily Grant",
            "type": "people"
          },
          {
            "name": "Eco Summit",
            "type": "organisation"
          },
          {
            "name": "City Media",
            "type": "organisation"
          }
        ]
      }
    ]
  },
\end{lstlisting}

\subsection{Aligning Story Elements}
\label{appendix:aligning-story-elements}
Before generating the text, we align story elements such as themes, entities, and time periods with each node in the graph. This alignment serves as a blueprint for the LLM to generate contextually relevant text. Here's an example of the alignment result:
\begin{lstlisting}[caption=JSON Map Example]
{
  "A1": {
    "Theme": "Economic Boom",
    "ThemeDescription": "The city experiences an unprecedented economic rise but faces a sudden ecological challenge before finding a sustainable economic solution.",
    "Entity": [
      "Downtown District",
      "Mayor Lina",
      "Economic Council"
    ],
    "Time": [
      "2023-01-01",
      "2023-05-02"
    ],
    "Prev": [],
    "StuctureInstruction": "The narrative structure is Arch, where The narrative begins with Theme A, then shifts to Theme B, and finally returns to Theme A. (eg. A1 -> B2 -> A3). The current chapter node is A1, which is the 1th chapter node in the narrative structure."
  },
  "B2": {
    "Theme": "Ecological Crisis",
    "ThemeDescription": "Due to the rapid economic growth, the city faces an environmental crisis, leading to debates on the city's future.",
    "Entity": [
      "Economic Council",
      "City Media",
      "Eco Summit",
      "GreenLife",
      "Riverstone Park"
    ],
    "Time": [
      "2023-07-01",
      "2023-07-31"
    ],
    "Prev": [
      "A1"
    ],
    "StuctureInstruction": "The narrative structure is Arch, where The narrative begins with Theme A, then shifts to Theme B, and finally returns to Theme A. (eg. A1 -> B2 -> A3). The current chapter node is B2, which is the 2th chapter node in the narrative structure."
  },
  "A3": {
    "Theme": "Economic Boom",
    "ThemeDescription": "The city experiences an unprecedented economic rise but faces a sudden ecological challenge before finding a sustainable economic solution.",
    "Entity": [
      "Downtown District",
      "Mayor Lina",
      "Economic Council",
      "Eco Summit"
    ],
    "Time": [
      "2023-08-31",
      "2023-12-31"
    ],
    "Prev": [
      "B2"
    ],
    "StuctureInstruction": "The narrative structure is Arch, where The narrative begins with Theme A, then shifts to Theme B, and finally returns to Theme A. (eg. A1 -> B2 -> A3). The current chapter node is A3, which is the 3th chapter node in the narrative structure."
  }
}
\end{lstlisting}

\subsection{Context Provider Prompt}
\label{appendix:context-prompt-template}
\begin{lstlisting}[caption=LLM Language Chain prompt template ]
Narrative Crafter is skilled in generating narrative contexts for various structures like Linear, Ladder, ShortFork, and others, while keeping the metadata ('name', 'description', 'structure', 'instructions', 'nodes', 'edges') unchanged. When modifying the 'context' section, which includes 'symbol', 'theme', 'description', 'time', and 'entities', I ensure that each theme is uniquely represented by a single letter symbol (A, B, C, etc.) and is explained in detail within its context. The theme symbol in the context section corresponds directly to the themes in the narrative structure. I avoid using the same theme symbol more than once in a single context to prevent confusion. I adhere to realistic naming conventions for people in entities, enhancing the narrative's authenticity. My responses strictly adhere to the JSON format, focusing on crafting contextually consistent and interconnected narrative contexts. In case of missing or unclear information, I will request clarification, aiming to deliver precise, coherent narrative contexts tailored to the specific narrative type and user input.
\end{lstlisting}

\subsection{Prompt Template}
\label{appendix:prompt-template}
To generate text for each node, we use a prompt template that includes the aligned story elements and the generated text from the previous nodes. Here's the prompt template:
\begin{lstlisting}[caption=LLM Language Chain prompt template ]
You are a senior reporter for The New York Times, your task is to write a fictional news report that continues from a previous one.
Your writing style is concise and clear, without rhetorical techniques, in plain english.
Avoid explicitly referencing the structure and themes; instead, integrate it subtly within the story.
The story should be written in a style suitable for CEFR Level A1 and each chapter approximately take 1 minutes to read (100 words).
The continuation chapter needs to be closely related to the previous one (intriguing and logically connected in narration).
The theme of the story must be strikingly prominent, you can not make up new themes except the only one given below.
All entities given below must be included in the story.
Chapter ID: {id}
Time Period: {time_period_str}
Themes: {theme}
Theme Context: {theme_context}
Entity: {', '.join(entity)}
Narrative Sturcture: {narrative_structure}
Previous context: {prev_content_str}
\end{lstlisting}

\subsection{Generated Story}
\label{appendix:generated-story}
Using the aligned story elements and the prompt template, the LLM generates the following story:
\begin{lstlisting}[caption=Textural Narrative]
Chapter A1 - "Downtown District Witnesses Unprecedented Economic Boom":
In a surprising turn of events, Downtown District is experiencing an economic boom. Mayor Lina announced yesterday that the district's economy has seen a significant rise in the past four months. The Economic Council attributes this growth to new businesses and increased tourism. However, Mayor Lina cautioned about an emerging ecological challenge. The city's infrastructure is under pressure due to the rapid economic growth. A sustainable solution is being sought to balance the boom and the environment.
Chapter B2 - "Eco Summit: A Turning Point for Downtown District":
This month, the city's economic boom was the main topic at the Eco Summit. The Economic Council and GreenLife, a local environmental group, led the discussions. They highlighted the growing ecological crisis in Riverstone Park. City Media reported that the park's wildlife is under threat due to increased tourism and infrastructure development. The Economic Council proposed a plan to balance economic growth and environmental preservation. The plan received mixed reviews, sparking debates on the city's future. The Eco Summit concluded with the call for sustainable solutions to the crisis.
Chapter A3 - "City Finds Sustainable Solution Amid Economic Boom":
As the year ends, the city finds a sustainable solution to the ecological crisis. The Economic Council, with Mayor Lina's support, introduces a new plan. It aims to balance the economic boom and environmental preservation. The plan includes limiting tourism and infrastructure development in the Downtown District. It also involves creating green jobs and promoting eco-friendly businesses. The city's economic growth continues, but with a new focus on sustainability. The plan receives positive feedback, showing a promising future for the city.
\end{lstlisting}
The generated story follows the "Arch" narrative structure, starting with the "Economic Boom" theme in A1, shifting to the "Ecological Crisis" theme in B2, and returning to the "Economic Boom" theme in A3 with a sustainable solution. The LLM incorporates the aligned story elements, maintains narrative coherence, and generates contextually relevant text for each node.

% \section{About Appendices}
% Refer to \cref{sec:appendices_inst} for instructions regarding appendices.

% \section{Troubleshooting}
% \label{appendix:troubleshooting}

% \subsection{ifpdf error}

% If you receive compilation errors along the lines of \texttt{Package ifpdf Error: Name clash, \textbackslash ifpdf is already defined} then please add a new line \verb|\let\ifpdf\relax| right after the \verb|\documentclass[journal]{vgtc}| call.
% Note that your error is due to packages you use that define \verb|\ifpdf| which is obsolete (the result is that \verb|\ifpdf| is defined twice); these packages should be changed to use \verb|ifpdf| package instead.

% \subsection{\texttt{pdfendlink} error}

% Occasionally (for some \LaTeX\ distributions) this hyper-linked bib\TeX\ style may lead to \textbf{compilation errors} (\texttt{pdfendlink ended up in different nesting level ...}) if a reference entry is broken across two pages (due to a bug in \verb|hyperref|).
% In this case, make sure you have the latest version of the \verb|hyperref| package (i.e.\ update your \LaTeX\ installation/packages) or, alternatively, revert back to \verb|\bibliographystyle{abbrv-doi}| (at the expense of removing hyperlinks from the bibliography) and try \verb|\bibliographystyle{abbrv-doi-hyperref}| again after some more editing.

\begin{algorithm}
  \caption{Generate Text from Structure Graph Data using LLM}
  \label{alg:generate_text}
  \begin{algorithmic}[1]
    \Require $narrative\_context$: List of narrative structures with context
    \Require $topic$: The topic for the story
    \Require $structure$: The selected narrative structure
    \Require $seed$: Optional seed for reproducibility
    \Procedure{GenerateStory}{$topic$, $structure$, $seed$}
    \State $narrative\_obj \gets$ find structure in $narrative\_context$ with name $structure$

    \State $alignment \gets$ \Call{AlignThemeEntity}{$narrative\_obj$, $seed$}
    \State $story \gets$ new Story with empty chapters
    \State $content\_cache \gets$ empty dictionary
    \For{$node$, $details$ in $alignment$}
    \Comment{Edge Traversal:}
    \State $prev\_content \gets$ concatenate content from $content\_cache$ for nodes in $details$["Prev"]
    \State $new\_chapter \gets$ \Call{GenerateChapter}{$node$, $details$, $prev\_content$}
    \State add $new\_chapter$ to $story.chapters$
    \State $content\_cache$[$node$] $\gets$ $new\_chapter.content$
    \EndFor
    \State \Return $story$
    \EndProcedure

    \Procedure{AlignThemeEntity}{$narrative\_obj$, $seed$}
    \State $alignment \gets$ empty dictionary
    \For{$node$ in $narrative\_obj.nodes$}
    \State $node\_context \gets$ find context in $narrative\_obj.context$ with symbol $node.symbol$
    \State $entities \gets$ sample entities from $node\_context.entities$
    \State $time\_period \gets$ generate time period for $node$
    \State $alignment$[$node$] $\gets$ {theme: $node\_context.theme$, entities: $entities$, time: $time\_period$}
    \EndFor
    \State \Return $alignment$
    \EndProcedure

    \Procedure{GenerateChapter}{$node$, $details$, $prev\_content$}
    \Comment{Node Expansion:}
    \State create $prompt\_template$ with $node$, $details$, $prev\_content$
    \State $model \gets$ OpenAI model with chosen parameters
    \State $output \gets$ $model(prompt\_template)$
    \State \Return $output$
    \EndProcedure
  \end{algorithmic}
\end{algorithm}

\begin{algorithm}
  \caption{Map Themes and Entities to Narrative Structure}
  \begin{algorithmic}[1]
    \If{seed is provided}
    \State Initialize random generator with seed
    \EndIf
    \State PreCalculateTimePeriods for narrativeObj
    \State ProcessNodesInOrder respecting parent-child relationships in narrativeObj
    \State Initialize an empty map for nodes
    \For{each node in processed order}
    \State Determine the node's symbol and fetch its narrative context
    \State Randomly sample entities for the node from its context
    \If{the node has parents}
    \State Ensure at least one entity from any of the parent nodes is included
    \EndIf
    \State Construct node map with:
    \begin{itemize}
      \item Theme from narrative context
      \item Randomly selected entities
      \item Time period calculated earlier
      \item Information about parent nodes
      \item Instructions or descriptions specific to the node's role in the structure
    \end{itemize}
    \EndFor
    \State \Return the constructed map for all nodes
  \end{algorithmic}
\end{algorithm}

\subsubsection*{Two TTNG Views}
\begin{figure*}[htp]
  \centering
  \includegraphics[width=\textwidth]{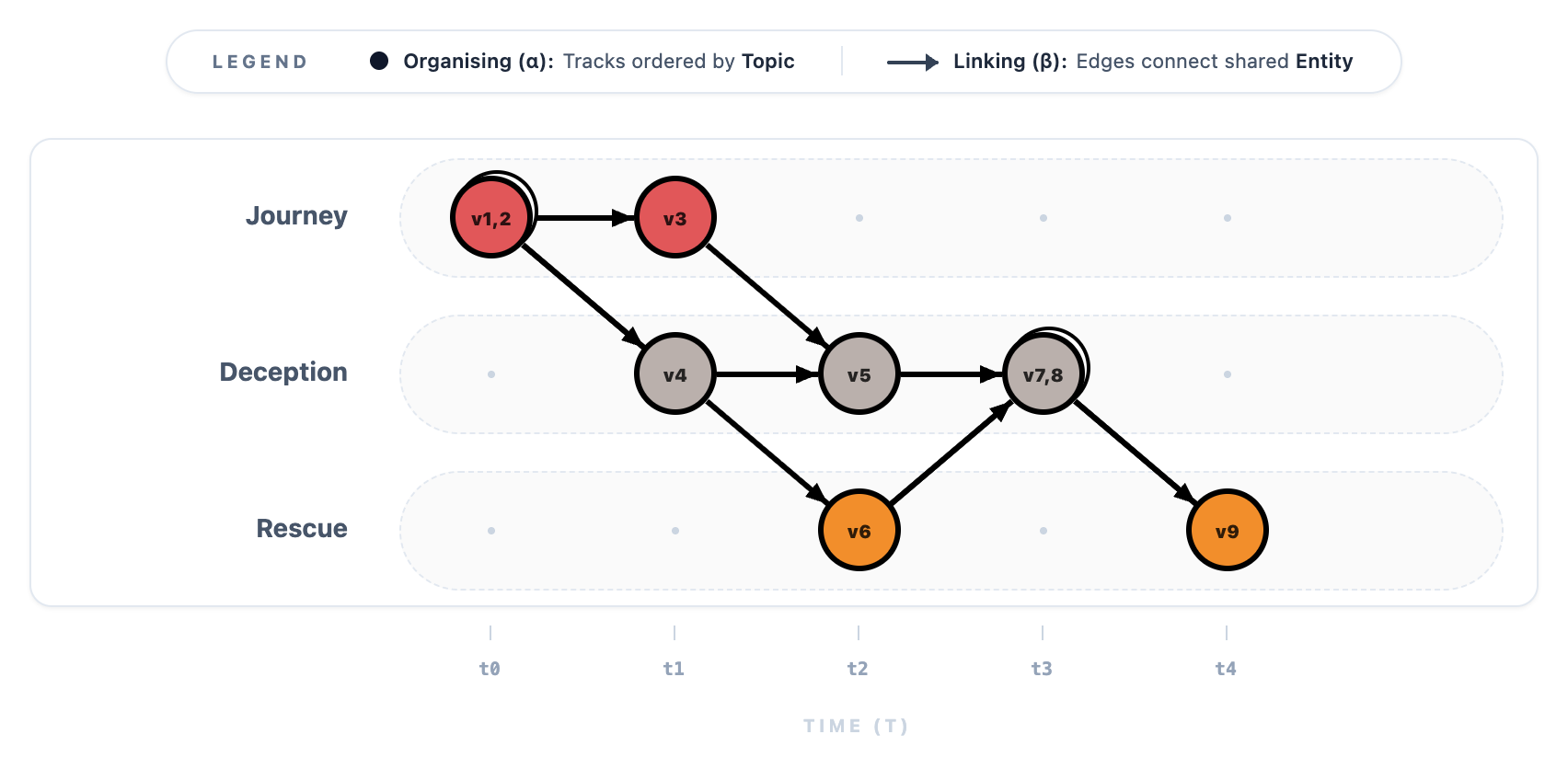}
  \smallskip
  \includegraphics[width=\textwidth]{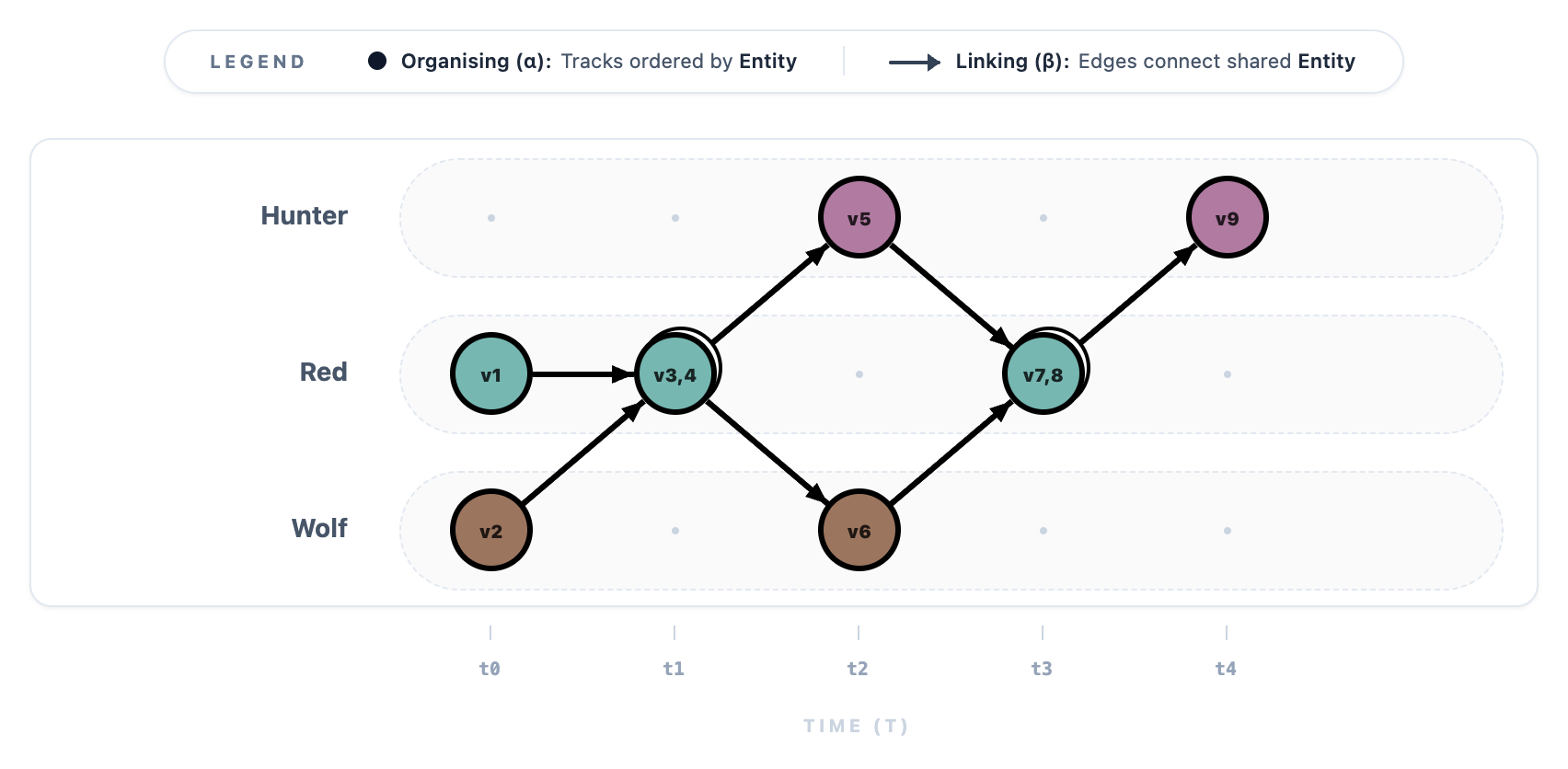}
  \caption{Two TTNG views for the same Little Red Riding Hood statements. Top: Topic-organised view ($\alpha = Topic$), showing Journey $\rightarrow$ Deception $\rightarrow$ Rescue. Bottom: Entity-organised view ($\alpha = Entity$), emphasising character interactions.}
  \label{fig:ttng_red_views}
\end{figure*}

\section{TTNG Worked Example and Motif Formalisation}
\label{appendix:ttng-example}

This appendix has two goals: (1) provide a worked TTNG example using \emph{Little Red Riding Hood}, and (2) formalise motif identification through constrained enumeration.

\subsection{Worked Example: Little Red Riding Hood}

\subsubsection*{Statements and Attributes}
\begin{table}[H]
  \centering
  \begin{tabularx}{\linewidth}{@{}l l l X@{}}
    \textbf{ID} & \textbf{Entities} & \textbf{Topic} & \textbf{Factual statement} \\
    \hline
    \textbf{V1} & Red & Journey & Red departs her home for Grandma's cottage carrying a basket of food through the woods. \\
    \rowcolor[HTML]{EFEFEF}
    \textbf{V2} & Wolf & Journey & The Wolf watches Red from the bushes near the dark forest entrance to plan his hunt. \\
    \textbf{V3} & Red, Wolf & Journey & Red stops to chat with the Wolf on the wooded path because he seems friendly. \\
    \rowcolor[HTML]{EFEFEF}
    \textbf{V4} & Wolf, Red & Deception & The Wolf encourages Red to pick wild flowers to delay the girl from reaching the cottage. \\
    \textbf{V5} & Wolf & Deception & The Wolf eats Grandma and dons her nightcap at the cottage to trick the girl. \\
    \rowcolor[HTML]{EFEFEF}
    \textbf{V6} & Hunter, Wolf & Rescue & The Hunter tracks the Wolf across the forest floor after hearing a suspicious disturbance. \\
    \textbf{V7} & Wolf, Red & Deception & The Wolf lunges at Red to eat her inside Grandma's quiet bedroom after revealing himself. \\
    \rowcolor[HTML]{EFEFEF}
    \textbf{V8} & Red, Wolf & Deception & Red enters the cottage and remarks on the Wolf's large ears while standing by the bed. \\
    \textbf{V9} & Hunter, Red, Wolf & Rescue & The Hunter kills the Wolf to rescue Red and Grandma from danger inside the house. \\
  \end{tabularx}
  \caption{Factual statements and attributes used in the Little Red Riding Hood TTNG example.}
  \label{tab:ttng_example_statements}
\end{table}

\Cref{fig:ttng_red_views} visualises these same nine units using two different \textbf{Organising Rules}. In TTNG, the vertical position of a node is determined by its track assignment, which is derived from a specific attribute.

\begin{itemize}
    \item \textbf{Topic View (Top):} Where $\alpha$ = Topic. The tracks represent abstract themes: \emph{Journey}, \emph{Deception}, and \emph{Rescue}. The narrative structure reveals thematic progression: it begins in the Journey track, shifts into the conflict of Deception, and resolves in the Rescue track. This view emphasises the \emph{logic of the plot}.
    \item \textbf{Entity View (Bottom):} Where $\alpha$ = Entity. The tracks represent concrete characters: \emph{Hunter}, \emph{Red}, and \emph{Wolf}. The structure is driven by interaction. The Wolf track acts as a central connector, interacting first with Red (during the deception phase) and then with the Hunter (during the rescue). This view emphasises \emph{character agency and interaction}.
\end{itemize}

By changing the Organising Rule, we change graph topology (the shape of the story) without changing temporal order or factual content. This captures the intuition that a single story can be ``told'' in multiple structural ways depending on focus.

Having demonstrated how a single set of facts can be viewed as different structures depending on the lens applied, we now turn to formalising the atomic units, or motifs, that make up these narrative structures.

\subsection{Narrative Motif Identification}
\label{appendix:narrative_motif_identification}

In graph theory, motifs typically refer to subgraphs that recur at a statistically significant rate compared to random networks.
In the context of TTNG, we use the term \emph{narrative motif} slightly differently: to denote the fundamental structural building blocks of narrative, analogous to sequence motifs in biology or narratemes in folklore studies. However, unlike narratemes, which classify \emph{semantic content} (e.g., ``hero leaves home''), our narrative motifs classify \emph{topological structure} (how tracks connect), regardless of the specific events.
Identifying these structural units allows us to decompose complex narrative graphs into smaller, interpretable components for evaluation.

\subsubsection*{Algorithmic Enumeration}
Given the constraints of the TTNG model, we can algorithmically enumerate all valid graph shapes within a narrative space of size $m \times n$.
To understand this, imagine a simple grid where rows are tracks and columns are time steps. We want to find every possible way to draw a valid story on this grid. Instead of manually sketching them, we use a mathematical approach to systematically generate every valid combination.

Given an $m \times n$ matrix $M$, where each cell $M_{i,j} \in \{0, 1\}$ represents the presence of a node at track $i$ and time $j$:

\[
M = \begin{bmatrix}
    M_{0,0} & M_{0,1} & \dots & M_{0,n-1} \\
    M_{1,0} & M_{1,1} & \dots & M_{1,n-1} \\
    \vdots & \vdots & \ddots & \vdots \\
    M_{m-1,0} & M_{m-1,1} & \dots & M_{m-1,n-1} \\
\end{bmatrix}
\]

\paragraph{Constraints for Valid Narratives.}
To represent a valid, connected narrative sequence, the matrix must satisfy three conditions:

\begin{enumerate}
  \item \textbf{Exact Number of Nodes}: The graph must contain exactly $n$ story units.
  \[
  \sum_{i=0}^{m-1} \sum_{j=0}^{n-1} M_{i,j} = n
  \]

  \item \textbf{Minimum Span}: The narrative must span at least two distinct time steps.
  \[
  \sum_{k=0}^{n-1} \mathbb{I}\left( \sum_{i=0}^{m-1} M_{i,k} > 0 \right) \geq 2
  \]
  \noindent where $\mathbb{I}(\cdot)$ is the indicator function, returning 1 if the condition is true (column $k$ is not empty) and 0 otherwise.

  \item \textbf{Canonical Form (No Isolated Timeframes)}: To avoid counting time-shifted duplicates (e.g., a pattern occurring at $t=1$ vs $t=2$), valid matrices must have no empty columns preceding non-empty ones.
  \[
  \forall k \in [0, n-2]: \quad \left( \sum_{i=0}^{m-1} M_{i,k} = 0 \right) \implies \left( \sum_{i=0}^{m-1} M_{i,k+1} = 0 \right)
  \]
\end{enumerate}

In simple terms, these conditions ensure that every valid story graph represents a coherent sequence where steps are not skipped, no timeframes are wasted, and the narrative has a definite beginning and end.

\paragraph{Objective.}
Our goal is to find all distinct matrices $M$ satisfying these constraints, where matrices are considered identical if they are row-equivalent (i.e., can be transformed into each other by swapping tracks).

\paragraph{Recursive Construction of Motifs.}
Let $f(m,n,X)$ be the set of valid matrices of size $m \times n$ with $X$ nodes.
For any matrix $A \in f(m,n,X)$, removing a specific node yields a matrix $A'$ that, after left-packing to satisfy the canonical-form constraint, belongs to $f(m,n,X-1)$.
Specifically:
\begin{enumerate}
    \item \textbf{Node Count:} Removal reduces the total count to $X-1$.
    \item \textbf{Span:} Since $A$ had at least two active columns, removing a single node typically leaves at least two active columns. For all $X \ge 3$, this property holds.
    \item \textbf{Canonical Form:} If removal creates an empty column $k$, we shift all subsequent columns $k+1 \dots n$ to the left (to $k \dots n-1$), restoring validity.
\end{enumerate}

\noindent\textbf{Conclusion:} This inductive property implies that any valid narrative graph of size $X$ is reachable from a smaller graph of size $X-1$. Complex narratives can therefore be analysed as sequences of node additions to basic 3-node patterns.
This justifies our focus on enumerating and evaluating the base case: \textbf{3-node narrative motifs}. Two-node motifs are trivial, while three-node motifs are the minimal complexity required to distinguish branching, merging, and turning.